\documentclass[useAMS,usenatbib]{mn2e}
\usepackage{graphicx}

\title[SN 1995N: probing the wind/ejecta interaction]
{Simultaneous {\it XMM-Newton} and ESO VLT observations of SN 1995N:
probing the wind/ejecta interaction
\thanks{Based on observations collected at ESO (programs
71.D-0265(A) and 71.D-0265(B)). Based on observations obtained with
{\it XMM-Newton} (Obs. ID 0149620201), an ESA science mission with
instruments and contributions directly funded by ESA Member States and
NASA.}}

\author[L. Zampieri, P. Mucciarelli, A. Pastorello, M. Turatto, E. Cappellaro, S. Benetti]
{L. Zampieri$^{1}$\thanks{E-mail: zampieri@pd.astro.it},
P. Mucciarelli$^{1,2}$\thanks{E-mail: mucciarelli@pd.astro.it},
A. Pastorello$^{3}$, M. Turatto$^{1}$, E. Cappellaro$^{4}$, \and S. Benetti$^{1}$\\
$^1$INAF--Osservatorio Astronomico di Padova, Vicolo dell'Osservatorio 5,
I-35122 Padova, Italy\\
$^2$Dipartimento di Astronomia, Universit\`a di Padova, Vicolo dell'Osservatorio 2, I-35122 Padova, Italy\\
$^3$Max Planck Institut f\"ur Astrophysik, Karl-Schwarzschild Strasse 1, 85741 Garching bei M\"unchen, Germany\\
$^4$INAF--Osservatorio Astronomico di Capodimonte, Via Moiariello 16,
I-80131 Napoli, Italy}

\begin{document}

\date{Accepted ... Received ...; in original form ...}

\pagerange{\pageref{firstpage}--\pageref{lastpage}} \pubyear{2004}

\maketitle

\label{firstpage}

\begin{abstract}
We present the results of the first {\it XMM-Newton} observation of
the interacting type IIn supernova 1995N, performed in July 2003.  We
find that the 0.2--10.0 keV unabsorbed flux dropped at a value of
$\simeq 1.8 \times 10^{-13}$ erg cm$^{-2}$ s$^{-1}$, almost one order
of magnitude lower than that of a previous {\it ASCA} observation of
January 1998. From all the available X-ray measurements, an
interesting scenario emerges where the X-ray light emission may be
produced by a two-phase (clumpy/smooth) CSM. The X-ray spectral
analysis shows statistically significant evidence for the presence of
two distinct components, that can be modeled with emission from
optically thin, thermal plasmas at different temperatures. The
exponent of the ejecta density distribution inferred from these
temperatures is $n\simeq 6.4$. From the fluxes of the two spectral
components we derive an estimate of the mass loss rate of the
supernova progenitor, ${\dot M} \sim 2 \times 10^{-4} M_\odot \, {\rm
yr}^{-1}$, at the upper end of the interval exhibited by red
super-giants. Coordinated optical and infrared observations allow us
to reconstruct the simultaneous infrared to X-ray flux distribution of
SN 1995N.  We find that, at $\sim$ 9 years after explosion, the direct
X-ray thermal emission due to the wind/ejecta interaction is $\sim 5$
times larger than the total reprocessed IR/optical flux.
\end{abstract}


\begin{table*}
\caption{Log of the {\it XMM-Newton} and VLT observations of SN 1995N}
\begin{flushleft}
\begin{tabular}{llllll}
\hline
Start Date      &MJD    &Observatory & Instrument            & Range/Filter/Grism
&Program/Obs. ID\\
\hline
2003-07-27      &52847.2  &{\it XMM-Newton} & EPIC pn+MOS, OM & 0.2--10 keV, $ubv \, UVW1 \, UVW2$ &0149620201  \\
2003-07-30      &52850.0  &VLT & FORS2              &$BVRI$, grism 300V+GG435   &71.D-0265(A)\\
2003-07-30      &52850.1  &VLT & ISAAC              &$JHK$
&71.D-0265(B)\\
\hline
\end{tabular}
\label{log}
\end{flushleft}
\end{table*}

\section{Introduction}
\label{sec1}

Among all core-collapse supernovae, the particular subclass of Type
IIn supernovae shows prominent multi-component hydrogen emission lines
in their optical spectra. Type IIn supernovae have been termed also
compact supernova remnants or Seyfert 1 impostors, because of the
peculiar profiles of the emission lines and the similarity of their
spectra to those of the Broad Line Region in AGNs. These lines are
thought to originate from the reprocessing of radiation generated by a
violent collision of the supernova ejecta with a dense surrounding gas
released by the progenitor star in a previous evolutionary stage. The
interaction generates forward and reverse shock waves which bound the
shocked wind and ejecta. This mechanism appears to be the
non-relativistic analogue of the blast wave interaction occurring in
$\gamma$-ray burst afterglows. The pressure and temperature behind the
shocks are sufficiently high that the post-shock ejecta and
circumstellar material (CSM) may become powerful X-ray emitters
\citep{cf94,cf01}. At the same time synchrotron radiation is generated
by electrons accelerated up to relativistic energies at the shock
front.

In the ``standard'' model of X-ray emission from circumstellar
interaction the forward shock produces a hot shell ($\sim 10^9$ K),
while the reverse shock produces a denser, cooler ($\sim 10^7$ K)
shell with much higher emission measure, from which the observed soft
X-ray emission arises \citep{cf94}. The post-shock material can be a
more efficient radiator if the explosion occurs in a dense
circumstellar shell with density $\sim 10^7$ cm$^{-3}$
\citep{terle92}. Alternatively, if the progenitor stellar wind is
clumpy, the interaction of the forward shock with clumps of gas can
give rise to cooler X-ray emission, with a temperature lower than that
expected from interaction with a smooth wind \citep{chugai93}.

Only $\sim$ 20 supernovae have been detected at X-ray energies. Recent
detections include SN 1988Z \citep{aretxaga99}, SN 1995N
\citep{fox00}, SN 1998bw \citep{iwa98}, SN 1999em and SN
1999gi \citep{schlegel01}. Although soon after the explosion the X-ray
emission may arise from Compton down-scattered $\gamma$-ray photons
produced in the radioactive decays of $^{56}$Ni and $^{56}$Co (see
e.g. the case of SN 1987A; \citealt{itoh87}), in most cases evidence
has been gathered in favour of the circumstellar interaction origin of
the observed radiation. SN 1988Z and SN 1995N are Type IIn supernovae
with a very large inferred X-ray luminosity ($L_X \sim 10^{41}$ erg
s$^{-1}$). SN 1999em and SN 1999gi are ``regular'' Type IIP (plateau)
supernovae whose X-ray emission is about two orders of magnitude lower
than that of Type IIn supernovae. This is usually attributed to the
fact that in these supernovae the CSM is less dense and hence the
ejecta interaction less strong. On the other hand, the powerful X-ray
emission of the ``hypernova'' SN 1998bw is probably originated by
synchrotron radiation produced by non-thermal electrons accelerated at
the shock front.

Several Type IIn supernovae show also an extraordinarily large
late-time infrared (IR) emission whose origin is still debated. The
data are interpreted as thermal emission from dust, but its origin and
the mechanism through which it is heated at the observed temperatures
($\sim$ 1000 K) does not appear to be established
\citep{gerardy02}. Several possibilities have been discussed,
including dust forming in the ejecta, pre-existing dusty circumstellar
medium heated by the supernova shock (or a precursor), IR echo powered
by emission from shock interaction with the circumstellar medium
\citep{pozzo04}.  For this type of supernovae multi-wavelength
observations may provide crucial information about the mass-loss
history in the late evolutionary stages of the progenitor star, the
distribution of the ejecta and circumstellar wind (CSW), and the
energy of the explosion.

Of special interest in this context is the case of SN 1995N,
discovered in May 1995 \citep{pa95}. Supernova 1995N is hosted in the
irregular galaxy complex known as Arp 261, at a distance of 24 Mpc,
whose center is occupied by the (IB(s)m pec) peculiar galaxy
MCG-02-38-017. The epoch of explosion is not known but may be
estimated to be about 10 months before the optical discovery
\citep{benetti95}. \cite{fran02} assumed an explosion date of 1994
July 4, about 2 years before the first X-ray observation. Throughout
this paper we adopt this assumption for the explosion epoch. SN 1995N
has been extensively observed in the optical
\citep{fran02,pasto05}. From an analysis of the emission line profile
in the optical and UV spectra at epochs between 321 and 1799 days
after the explosion, \cite{fran02} found evidence of three distinct
velocity components for the gas, suggesting a clumpy circumstellar
medium or an aspherical distribution of the surrounding gas.


X-ray emission has been detected with ROSAT and ASCA at different
epochs. The first X-ray observation of SN 1995N was performed with
{\it ROSAT} HRI on 1996 July 23 (1.3 ks; \citealt{lewin96}), followed
by other two exposures taken on 1996 August 12 (17 ks) and on 1997
August 17 (19 ks; see \citealt{fox00}).  An {\it ASCA} observation was
performed on 1998 January 19 ($\sim$ 90 ks for SIS and GIS;
\citealt{fox00}).

The high X-ray luminosity ($L_X \sim 10^{40}-10^{41}$ erg s$^{-1}$)
places SN 1995N in a small group of Type IIn supernovae with strong
circumstellar interaction (along with SN 1978K, SN 1986J, SN 1988Z),
making it an ideal target for studying the X-ray spectral evolution
and light curve of Type IIn supernovae. The X-ray light curve from 2
to 3.5 years after explosion suggests that the CSM is distributed
inhomogeneously and that the average X-ray luminosity does not decline
significantly \citep{fox00}. In the model of \cite{cf94} a decline in
flux is expected if the forward/reverse shock luminosity is dominated
by free-free emission, whereas a radiative reverse shock front leads
to a roughly constant luminosity when line emission is dominant. Both
behaviours have been observed (in SN 1986J and SN 1978K,
respectively). \cite{fox00} found that the {\it ASCA} spectrum of SN
1995N is consistent with a simple bremsstrahlung model with absorption
($kT\simeq 9$ keV, $N_H \simeq 10^{21}$ cm$^{-2}$), although also a
power-law with photon index $\alpha\simeq 1.7$ and a slightly higher
column density provides an acceptable fit. The spectral fitting hints
towards the presence of spectral features near $\sim$ 1.8 keV,
appropriate for fluorescent silicon emission (Si K-shell lines).

In this paper we present the results of a {\it XMM-Newton} observation
of SN 1995N, performed in July 2003, along with those obtained from a
simultaneous optical/infrared observation carried out at ESO. The
paper is organized as follows. In \S \ref{sec2} we present our {\it
XMM}-Newton observation and the adopted X-ray data reduction
procedure. \S \ref{sec3} is devoted to the X-ray spectral analysis of
SN 1995N and two other field sources, while in \S \ref{sec4} we
present the OM images of the field. \S \ref{sec5} reports the X-ray
light curve of SN 1995N from all the available X-ray measurements.  In
\S \ref{sec6} we present the results from our coordinated optical/IR
observations of SN 1995N. Finally, results are discussed in \S
\ref{sec7} and conclusions summarized in \S \ref{sec8}.

\begin{figure*}
 \vbox{\vskip 12truecm}
 \caption{{\it XMM} EPIC image of SN 1995N taken on July
 27--28, 2003. The field is centered on the supernova and contains
 other two sources of comparable brightness, 2 and 7 (also labeled
 X-1 and X-3; see Table \ref{tabfield}). The latter is coincident with
 the nucleus of the galaxy MCG-02-38-017.
 [{\bf This Figure is available as a separate jpg file}].}  \label{fig1}
\end{figure*}

\begin{table*}
\caption{Field sources (in order of decreasing RA) in the {\it XMM} EPIC image of SN 1995N}
\begin{flushleft}
\begin{tabular}{lllllll}
\hline
Source number & Source name & RA [J2000] & DEC [J2000] & MOS count rate$^a$ & pn count rate$^a$ \\
& & (hh:mm:ss.sss) & ($^0$ : $'$ : $''$) & ($10^{-2}$ s$^{-1}$) & ($10^{-2}$ s$^{-1}$)\\
\hline
1     &     & 14:49:08.616  &  $-$10:18:58.75 & -- & -- \\
2     & X-1 & 14:49:15.088  &  $-$10:11:16.26 & 1.15$\pm$0.11 & 3.66$\pm$0.17 \\
3     &     & 14:49:16.410  &  $-$10:16:01.28 & -- & -- \\
4     &     & 14:49:23.522  &  $-$10:14:02.68 & -- & -- \\
5     & X-2/SN 1995N & 14:49:28.359  &  $-$10:10:14.52 & 0.96$\pm$0.07 & 3.3$\pm$0.16  \\
6     &     & 14:49:30.762  &  $-$10:10:33.77 & -- & -- \\
7     & X-3 & 14:49:32.720  &  $-$10:09:48.75 & 0.85$\pm$0.10 & --            \\
8     &     & 14:49:53.662  &  $-$10:06:22.50 & -- & -- \\
9     &     & 14:49:53.927  &  $-$10:15:16.25 & -- & -- \\
10    &     & 14:49:56.098  &  $-$09:59:41.23 & -- & -- \\
11    &     & 14:50:08.673  &  $-$10:11:45.12 & -- & -- \\
12    &     & 14:50:11.622  &  $-$10:13:23.64 & -- & -- \\
\hline
\end{tabular}

$^a$ Background subtracted

\label{tabfield}
\end{flushleft}
\end{table*}

\begin{table*}
\caption{Parameters of the joint fit of the {\it XMM} EPIC MOS and pn
observation of SN 1995N (0.2-10.0 keV).}
\begin{flushleft}
\begin{tabular}{lllll}
\hline
Model & $\frac{N_{H}}{10^{21} cm^{-2}}$ & Parameter(s) &
$\chi^2_{red}$(dof) & Flux$^a$\\
\hline
POWER-LAW       &1.9$_{-0.4}^{+0.4}$     &$\Gamma$=2.0$_{-0.2}^{+0.2}$
&1.18(58) & $2.08^{+0.71}_{-0.45}$ \\
MEKAL$^b$           &0.1$_{-0.3}^{+0.3}$     &$kT$=4.6$_{-0.9}^{+1.3}$ keV
&1.21(58) & $1.61^{+0.28}_{-0.26}$ \\
POWER-LAW+MEKAL$^b$ &1.4$_{-0.4}^{+0.4}$     &$\Gamma$=1.5$_{-0.2}^{+0.3}$
&0.78(56) & $1.87^{+1.05}_{-0.70}$ \\
                &                        &$kT$=0.8$_{-0.1}^{+0.1}$  keV  & \\
MEKAL$^b$+MEKAL$^b$     &1.3$_{-0.4}^{+0.5}$   &$kT_1$=0.8$_{-0.1}^{+0.1}$ keV
&0.76(56) & $1.76^{+0.17}_{-0.38}$ \\
                &                        &$kT_2$=9.4$_{-4.2}^{+22.6}$ keV& \\
\hline
\end{tabular}

$^a$ Unabsorbed 0.2-10 keV flux in units of $10^{-13}$ erg cm$^{-2}$
s$^{-1}$.

$^b$ Solar abundance (frozen).

\label{tab1}
\end{flushleft}
\end{table*}

\begin{figure*}
 \begin{center}
 \includegraphics[height=14truecm,angle=-90]{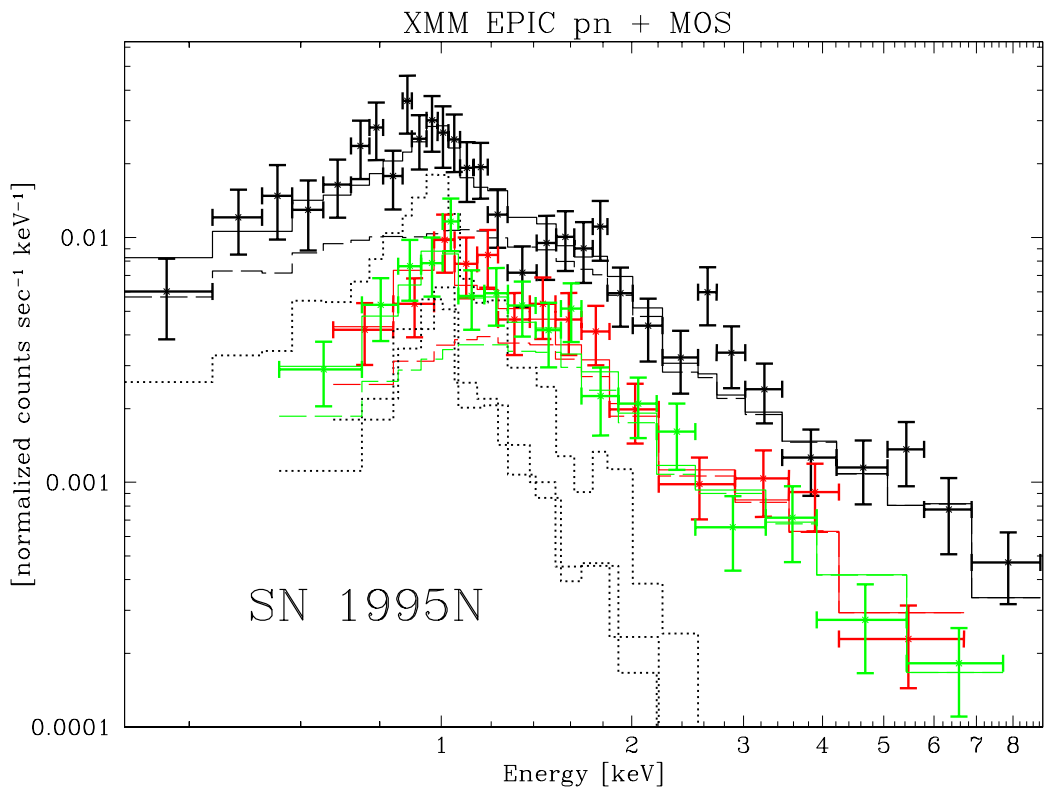} \end{center}
 \caption{{\it XMM} EPIC MOS and pn spectra of SN 1995N in the
 0.2-10.0 keV interval (pn above), along with the best fitting
 continuum model (solid lines) and the low/high temperature MEKAL
 components (dotted/dashed lines) for the three instruments.  } 
 \label{fig2}
\end{figure*}

\section{X-ray observation and data reduction}
\label{sec2}

We observed SN 1995N with {\it XMM-Newton} on July 27--28, 2003,
$\sim$ 9 years after the estimated date of explosion (Table
\ref{log}). The observation (ID 0149620201) lasted 72 ks, but was
heavily affected by solar flares. The three {\it XMM} EPIC cameras
(MOS1, MOS2 and pn) operated in Prime Full Window mode with the thin
filter. The EPIC MOS observation was split into two exposures of $\sim
5$ and $\sim 56$ ks, respectively, while only one uninterrupted EPIC
pn exposure of $\sim 64$ ks was taken.  The two RGS instruments show
insufficient counts to perform any statistically significant analysis.

Data screening, region selection and event extraction were performed
using the XMMSAS (v 5.4) software package. Data were filtered using
the good time intervals when the total off-source count rate above 10
keV $C$ was $< 0.5$ counts s$^{-1}$ for the EPIC MOS and $< 1$ counts
s$^{-1}$ for the EPIC pn, leaving $\sim$ 22 and 14 ks of useful data,
respectively. After filtering, the first, shorter EPIC MOS exposure
had a very small number of source photons ($\sim$ 25) and was then
discarded. Although the standard criterion for filtering MOS data
recommended by the {\it XMM} Science Operation Center
(XMM-SOC-CAL-TN-0018) is $C < 0.35$ counts s$^{-1}$, we decided to
take a slightly higher threshold ($C < 0.5$ counts s$^{-1}$) to have
an acceptable statistics in the MOS data. Any residual contamination
caused by solar flares should not affect seriously our results because
intense solar flares typically have $C \geq 10-20$ counts s$^{-1}$ and
EPIC MOS data have been analyzed jointly with the pn data (always
below the recommended threshold).

\begin{table*}
\caption{Parameters of the joint fit of the {\it XMM} EPIC MOS and pn
observation of field sources X-1 and X-3 (0.2-10.0 keV).}
\begin{flushleft}
\begin{tabular}{lllll}
\hline
Source & Model & $\frac{N_{H}}{10^{21} cm^{-2}}$ & Parameter(s) &
$\chi^2_{red}$(dof) \\
\hline
X-1 & POWER-LAW       &       0.8$_{-0.4}^{+0.3}$
&$\Gamma$=2.1$_{-0.1}^{+0.1}$      & 0.89(65)\\
    & BREMSSTHRALUNG  &       0.1$_{-0.1}^{+0.2}$
&$kT$=3.5$_{-0.8}^{+1.2}$  keV     & 0.98(65)\\
X-3 & POWER-LAW       &       2.8$_{-1.2}^{+0.9}$
&$\Gamma$=2.8$_{-0.5}^{+0.4}$      & 0.98(36)\\
    & BREMSSTHRALUNG  &       1.1$_{-0.5}^{+0.8}$
&$kT$=1.9$_{-0.6}^{+0.8}$  keV     & 1.01(36)\\
\hline
\end{tabular}

\label{tab1b}
\end{flushleft}
\end{table*}

A filtered {\it XMM} EPIC image of the field is shown in Figure
\ref{fig1}. A list of sources with peak counts larger than the average
counts of the adjacent background is reported in Table \ref{tabfield}.
The position of SN 1995N (RA=14h49m28s.359, DEC=$-10^0$10$'$14$''$.52
[J2000]) is within 0.6$''$ from the accurate radio position by Van Dyk
et al. (1996) (RA=14h49m28s.313, DEC=$-10^{0}$10$'$13$''$.92
[J2000]). Besides SN 1995N, other two sources appear to have
comparable brightness, a field object (labeled X-1) positionally
coincident with a cataloged star (GSC2 S231011112849) and the nucleus
of the galaxy MCG-02-38-017 (labeled X-3).

\section{X-ray spectral analysis}
\label{sec3}

\subsection{{\it XMM} EPIC spectrum of SN 1995N}

Source counts were extracted from a circular region of radius 20$''$
centered on the source position. Background counts were extracted from
a circular region of radius 40$''$, on the same CCD. The net source
count rate is reported in Table \ref{tabfield}. A total of $\sim$ 420
and $\sim$ 460 photons were collected from the EPIC MOS and pn
cameras, respectively.

{\it XMM} EPIC MOS and pn spectra were binned requiring at least 15
counts per bin. They are shown in Figure \ref{fig2}. Joint MOS and pn
spectral fits were performed in the 0.2-10.0 keV interval with XSPEC
(v. 11.2.0). An overall normalization constant was included to
minimize the effects of possible relative calibration uncertainties
among the different instruments (typical fractional difference
5-10\%). Despite the low statistics, the fit with single component
models is not fully satisfactory ($\chi^2_{red}
\simeq 1.2$; see Table \ref{tab1}). The best fit is obtained with a
dual MEKAL\footnote{The MEKAL model is the spectrum emitted by an
optically thin, thermal plasma.}  model, convolved with interstellar
absorption (see Figure \ref{fig2}). The best fitting column density of
the interstellar medium is $N_H = 1.3 \times 10^{21}$ cm$^{-2}$, while
the temperatures of the two MEKAL components are $kT\simeq 0.8$ and
9.5 keV (see Table \ref{tab1}). The improvement over a single
component model is significant at the 4.1 $\sigma$ level. We tried
also to perform spectral fits by varying the chemical composition of
the two MEKAL components with respect to solar. Increasing or
decreasing the chemical abundances up to $\sim$ 30\% in both
components gives spectral fits comparable to those obtained with solar
composition.

Taking the best-fitting spectral model (MEKAL+MEKAL), the
absorbed/unabsorbed flux in the 0.2-10.0 keV band is $F=1.46 \times
10^{-13}$/$1.76 \times 10^{-13}$ erg cm$^{-2}$ s$^{-1}$ (corresponding
to a luminosity L = $9.4 \times 10^{39}$ erg s$^{-1}$, at the distance
of the host galaxy $d=24$ Mpc), about an order of magnitude lower than
that of the {\it ASCA} observation of January 1998.

\subsection{{\it XMM} EPIC spectrum of sources X-1 and X-3}

The spectral analysis for sources X-1 and X-3 was performed as
described in the previous section. The count rates are reported in
Table \ref{tabfield}. Source X-3 falls in a gap in the EPIC pn image
and then only the MOS data were used. For X-1, joint MOS and pn
spectral fits were performed. In both cases spectral fits were
performed in the 0.2-10.0 keV interval with an overall normalization
constant. As shown in Table \ref{tab1b}, both an absorbed
bremssthralung and a power-law can satisfactorily reproduce the
observed spectrum, while a single blackbody component is ruled out by
the data ($\chi^2_{red}=2.4$ and 1.4, respectively). Assuming a
power-law spectrum with the parameters reported in Table \ref{tab1b},
the unabsorbed fluxes in the 0.2-10.0 keV band derived from the EPIC
MOS cameras are $F=2 \times 10^{-13}$ erg cm$^{-2}$ s$^{-1}$ and $F=3
\times 10^{-13}$ erg cm$^{-2}$ s$^{-1}$ for sources X-1 and X-3,
respectively. From the parameters of the X-ray spectral fit and the
existence of a likely counterpart identified with the GSC2 field star
S231011112849, we tentatively conclude that source X-1 may be an X-ray
binary in our Galaxy. The space density of such objects at this flux
level is not known. As mentioned in \S~\ref{sec2}, object X-3 is
spatially coincident with the nucleus of the galaxy MCG-02-38-017.
The best fitting power-law model gives parameters roughly consistent
with emission from a low-luminosity AGN.

\begin{figure*}
 \begin{center}
 \includegraphics[height=8truecm]{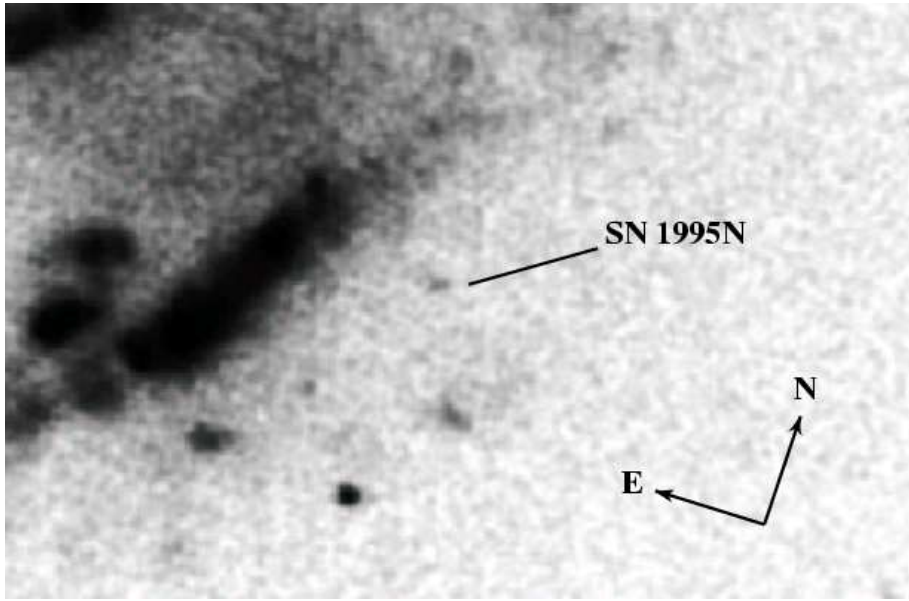}
 \end{center} \caption{OM image of the field around SN 1995N in the $UVW1$
  band (taken on July 27--28, 2003), showing that the supernova is well detected.}
 \label{fig4}
\end{figure*}

\begin{figure*}
 \begin{center}
 \includegraphics[height=8truecm]{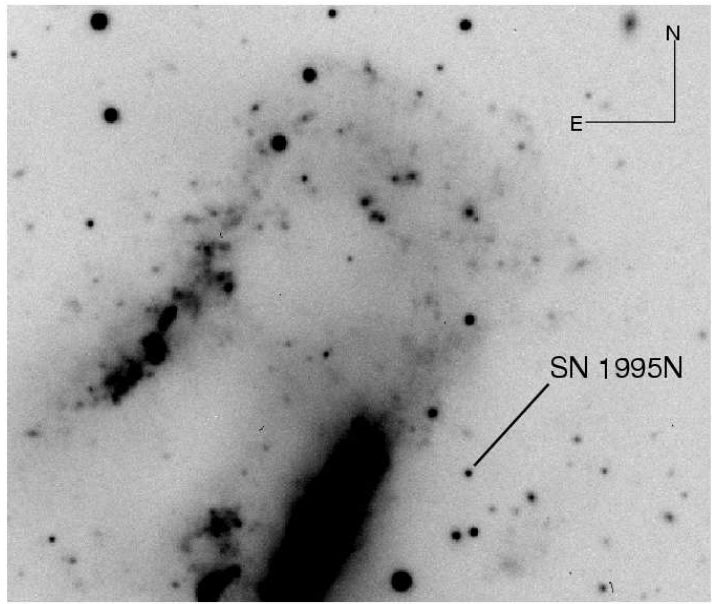}
 \end{center} \caption{VLT $R$-band image of the field around SN 1995N taken on July 30, 2003.}
 \label{fig5}
\end{figure*}

\begin{figure*}
 \begin{center}
 \includegraphics[height=16.5truecm,angle=-90]{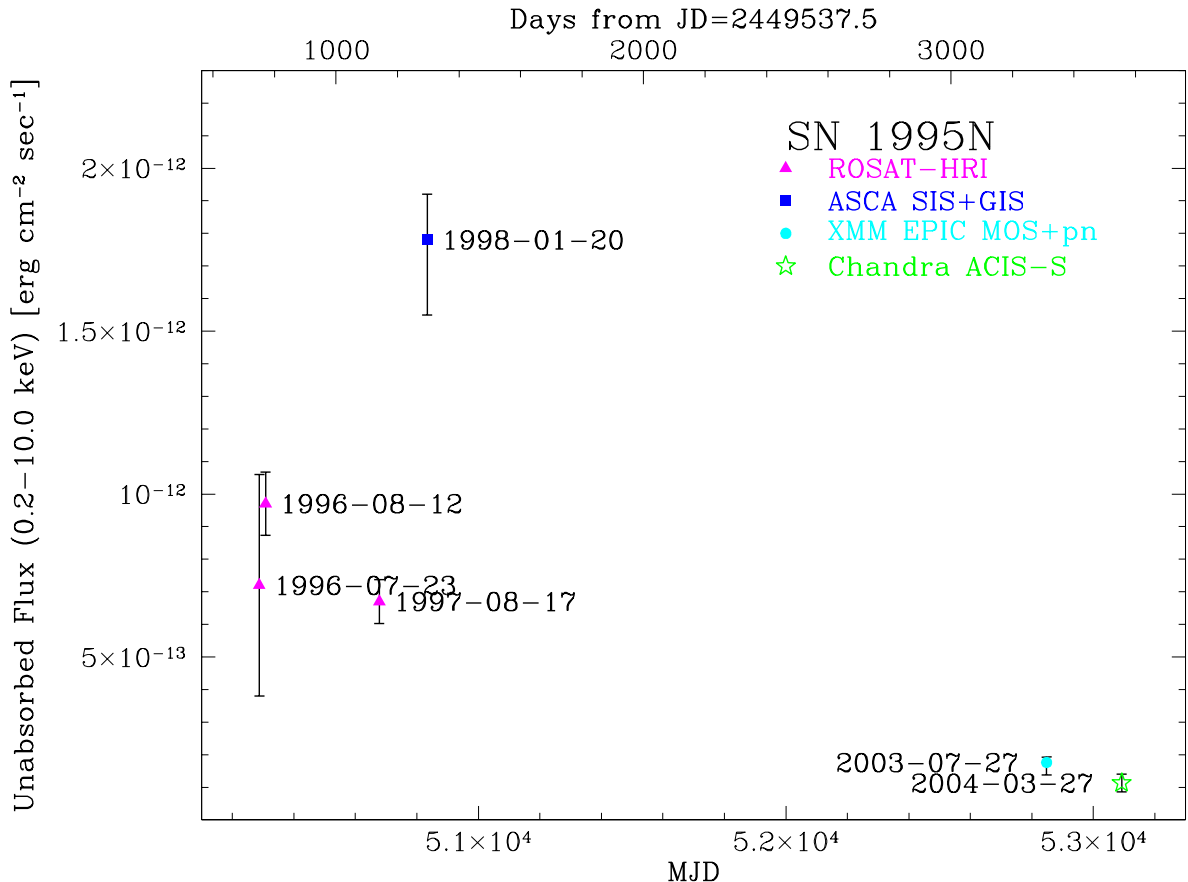} \end{center}
 \caption{0.2-10.0 keV unabsorbed lightcurve of SN 1995N. Fluxes were
 de-absorbed and, when necessary, extrapolated in the 0.2-10.0 keV
 interval. For the {\it ROSAT} HRI data we adopt the count rates from
 Fox et al. (2000), assuming a dual Raymond-Smith model with the same
 parameters reported in Table \ref{tab1}. For the {\it ASCA} GIS+SIS
 data we use to the values reported by Fox et al. (2000), after
 subtracting the contribution from source X-3. The {\it Chandra} point
 is from Chandra et al. (2005). Finally, for the {\it XMM} EPIC data,
 the fluxes were calculated using the best fitting spectrum with the
 parameters reported in Table \ref{tab1}.  The error on the {\it XMM}
 point is calculated propagating the uncertainty on the spectral
 parameters.}  \label{fig3}
\end{figure*}

\section{OM imaging}
\label{sec4}

During the {\it XMM-Newton} observation of SN 1995N a series of $\sim$
5000 s exposures of the field in the $v$, $b$, $u$, $UVW1$ and $UVW2$
bands was also taken with the OM\footnote{Note that the {\it XMM} OM
$u$, $b$, $v$ bands do not exactly coincide with the standard Johnson
$U$, $B$ and $V$ band-passes. For a definition of the effective
wavelengths of these filters see e.g. \cite{royer00} and
\cite{mason01}.}. The exposures (3 per band) were astrometrically
calibrated using standard IRAF tasks starting from the positions of
GSC2 ESO field stars. The intrinsic accuracy of this procedure is
0.5$''$. The calibrated images were then summed together. Nothing is
visible at the position of SN 1995N in the $u$, $b$, $v$ and $UVW2$
bands. Using the detection limit count rates of the single images and
the conversion factors reported in the XMM-SAS User's Guide (v. 3.0),
we estimate the following upper limits for the flux of SN 1995N:
$F_v<7.7 \times 10^{-15}$ erg cm$^{-2}$ s$^{-1}$, $F_b<1.1 \times
10^{-14}$ erg cm$^{-2}$ s$^{-1}$, $F_u<3.5
\times 10^{-15}$ erg cm$^{-2}$ s$^{-1}$, $F_{UVW2}<2.3 \times 10^{-14}$
erg cm$^{-2}$ s$^{-1}$.

A faint object is indeed detectable inside the X-ray error box of SN
1995N in the summed $UVW1$ image shown in Figure \ref{fig4}. The
object is at the limit of detectability in each single frame. The
position in the summed image is RA=14h49m28s.336,
DEC=$-10^0$10$'$13$''$.77, within 0.4$''$ from the radio position of
SN 1995N. Within the errors of the astrometric calibration, we then
identify this object with SN 1995N. The lack of catalogs of reference
stars in the $UVW1$ band prevent us from performing an accurate
photometric calibration in this band.  Thus, in order to estimate a
lower limit for the flux emitted by SN 1995N, we use the detection
limit count rate of a single image. Because SN 1995N is slightly above
threshold, we can safely assume $F_{UVW1}\ga 6 \times 10^{-15}$ erg
cm$^{-2}$ s$^{-1}$.

\section{X-ray light curve of SN 1995N}
\label{sec5}

Figure \ref{fig3} shows the 0.2-10 keV unabsorbed fluxes derived from
all the available X-ray observations of SN 1995N. The {\it ROSAT}
fluxes were derived assuming a dual Raymond-Smith model (similar to
the best fitting dual MEKAL model) with the same parameters reported
in Table \ref{tab1}.
For the {\it ASCA} fluxes, the original measurement from
\cite{fox00}, averaged between the SIS and GIS instruments and extrapolated to the
0.2-10 keV interval, is $F=1.8 \times 10^{-12}$ erg cm$^{-2}$
s$^{-1}$.  It should be noted, however, that in our {\it XMM} image
two field sources (6 and 7) are sufficiently close to the supernova to
be unresolved in the previous {\it ASCA} observation.  In particular,
the flux emitted by one of these sources (7 or X-3) is not negligible
in comparison with that of SN 1995N. Assuming that it remained
constant and subtracting it from that of the supernova, we find that
the actual {\it ASCA} flux of SN 1995N should have been $F= 1.5
\times 10^{-12}$ erg cm$^{-2}$ s$^{-1}$. This is the value reported in
Figure \ref{fig3}.  For {\it XMM}, the fluxes are the average between
the different instruments (MOS1, MOS2, pn, respectively). The {\it
XMM} value refers to the best fitting MEKAL+MEKAL model ($F=1.76
\times 10^{-13}$ erg cm$^{-2}$ s$^{-1}$; see Table
\ref{tab1}). Finally, we report also a recent measurement obtained
with {\it Chandra} \citep{chan05}, extrapolated to the 0.2-10 keV
interval.

It is worth noting in Figure \ref{fig3} the large decrease in the
X-ray flux between the {\it ASCA} and {\it XMM} observations,
indicating that the supernova has started its X-ray decline. This
conclusion is independent of the different spectral models used to
calculate the {\it XMM} flux (see Table \ref{tab1}), and is consistent
with the recent {\it Chandra} measurement.

\begin{figure*}
 \begin{center}
 \includegraphics[height=15.0truecm,angle=-90]{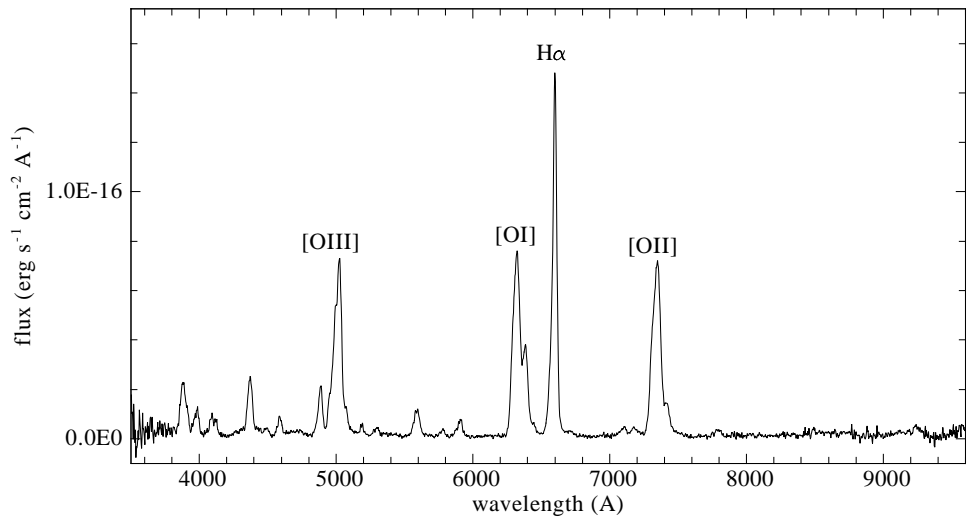}
 \end{center} \caption{VLT/FORS2 spectrum of SN 1995N (3500-9600 A)
 taken on July 30, 2003 (grism 200V+GG435).}
 \label{fig5spe}
\end{figure*}

\section{Coordinated optical and Infrared observations of SN 1995N}
\label{sec6}

Coordinated optical and IR observations of SN 1995N were performed
with the ESO VLT telescopes (VLT+FORS2, VLT+ISAAC) on July 30, 2003
(Table \ref{log}). Two optical spectra of 1460 s each were obtained
with UT4 equipped with FORS2, using the grism 200V and the separation
order filter GG435. The spectral resolution is about 10 A and the
wavelength range is 4200-9600 A. All data were reduced in the IRAF
environment, either using IRAF tasks or specific procedures developed
by our group. The two spectra were combined together and calibrated in
wavelength using comparison spectra of He-Ar and He-Ne lamps. The flux
calibration was performed using spectrophotometric standards. The
estimated error on the flux is $\sim 20\%$. The VLT spectrum is shown
in Figure \ref{fig5spe}.
Photometry in the $BVRIJHK$ bands was performed on the VLT images,
using a PSF-fitting technique. Photometric errors were estimated
placing artifical stars with the same magnitude and profile as the SN,
at positions close (few arcsecs) to that of the SN, and then computing
the deviations of the measured artificial stars magnitudes. All the
photometric measurements are reported in Table \ref{tab2}. The $R$
band VLT image is shown in Figure \ref{fig5}.

After $\sim 9$ years SN 1995N is still well detected in all optical/IR
bands. During the 1998-2003 time interval, the decline in the $K$ and
$V$ band magnitudes has been $\sim 3.8$ and $\sim$ 2.2 mag,
corresponding to a flux decrement of a factor $\sim$ 30 and $\sim$ 8,
respectively.  This is rather slow if compared to the typical
radioactive decay decline of Type II SNe ($\sim 1$ mag/100 days for
$^{56}$Co), confirming that the IR/optical emission is powered by the
interaction of the ejecta with the CSM (with radioactive decay of
heavy isotopes playing only a minor role). A comparison of the decline
rate of SN 1995N in the X-ray and $K$ bands shows that, between 1998
and 2003 (epochs of the {\it ASCA} and {\it XMM} observations), the
X-ray flux has decreased by a factor $\la$ 10, while the $K$ band
emission has dropped much more (a factor $\sim$ 30). In 2003 the X-ray
luminosity exceeds that in the $K$ band by approximately one order of
magnitude (see Table \ref{tab3}). Thus, in 1998, $F_X/F_K \sim 3$.


The optical spectrum is dominated by the H balmer lines, [O III]
4959-5007\AA, [O I] 6300-6364 \AA, [O II] 7320-7330 \AA (see Figure
\ref{fig5spe}). In addition, weak emission features of He I are
visible. The total flux of H$\alpha$ decreases by a factor 60 between
the first spectrum of Fransson et al. 2002 (1996 June 22) and our VLT
spectrum (2003 July 30). In the same period, the flux of H$\beta$
decreases by a factor 12 (i.e. the Balmer decrement reduces by a
factor 5). Similar to H$\beta$ is the evolution of the total flux for
the He I and [O III] 4969-5007\AA lines. However, over the same period
the fluxes of the [O I] and [O II] doublets show very small changes,
remaining roughly constant.

The H emission lines in our VLT spectrum of July 30, 2003, show an
asymmetric profile, with two different components: a broader component
with FWHM = 2670 km s$^{-1}$ and a narrower one with FWHM = 1270 km
s$^{-1}$. As noted by \cite{fran02}, the H$\alpha$ profile has a
significant evolution with time, showing a deficit of flux in the red
wing compared to the blue wing at the same velocity. Moreover, the
broader emission component seems to peak at bluer wavelengths. This
may be explained by the contribution of dust formed into the ejecta or
pre-existing in the CSM (e.g. \citealt{pozzo04,gerardy02}).

\begin{table*}
\caption{Optical and infrared photometry of SN 1995N from the ESO VLT observation of July 30, 2003}
\begin{flushleft}
\begin{tabular}{lllllllllllllllll}
\hline
date & MJD & $B$ & $V$ & $R$ & $I$ & $J$ & $H$ & $K$ & instr. \\
\hline
30/07/03 & 52850.0 & 22.62 & 21.84 & 20.79 & 21.00 & -- & -- & -- & 1 \\
30/07/03 & 52850.1 & -- & -- & -- & -- & 20.96 & 19.75 & 18.01 & 2 \\
\hline
\end{tabular}

1 VLT+FORS2; 2 VLT+ISAAC

\label{tab2}
\end{flushleft}
\end{table*}

\begin{table*}
\caption{Infrared through X-ray flux of SN 1995N from our multi-wavelength observations of July 2003}
\begin{flushleft}
\begin{tabular}{lllllllllll}
\hline
$F_X^{a,b}$ & $F_{UVW2}^{a,c}$ & $F_{UVW1}^{a,d}$ & $F_u^{a,e}$ & $F_B^a$
& $F_V^a$ & $F_R^a$ & $F_I^a$ & $F_J^a$ & $F_H^a$ & $F_K^a$ \\
\hline
$1.76^{+0.17}_{-0.38}$ & $<$ 0.23 & $\ga$ 0.060 & $<$ 0.035 &
0.058$^f$ & 0.061$^f$ & 0.13$^f$ & 0.059$^f$ & 0.026$^f$ & 0.059$^f$ &
0.13$^f$ \\
\hline
\end{tabular}

$^a$ Fluxes are in units of $10^{-13}$ erg cm$^2$ s$^{-1}$

$^b$ [0.2-10] keV band

$^c$ Effective wavelength of the $UVW2$ filter: 2070 A \citep{royer00}

$^d$ Effective wavelength of the $UVW1$ filter: 2905 A \citep{royer00}

$^e$ Effective wavelength of the $u$ filter: 3472 A \citep{royer00}

$^f$ Fractional error $\simeq$ 1\%

\label{tab3}
\end{flushleft}
\end{table*}

\begin{figure*}
 \begin{center}
 \includegraphics[height=14.0truecm,angle=-90]{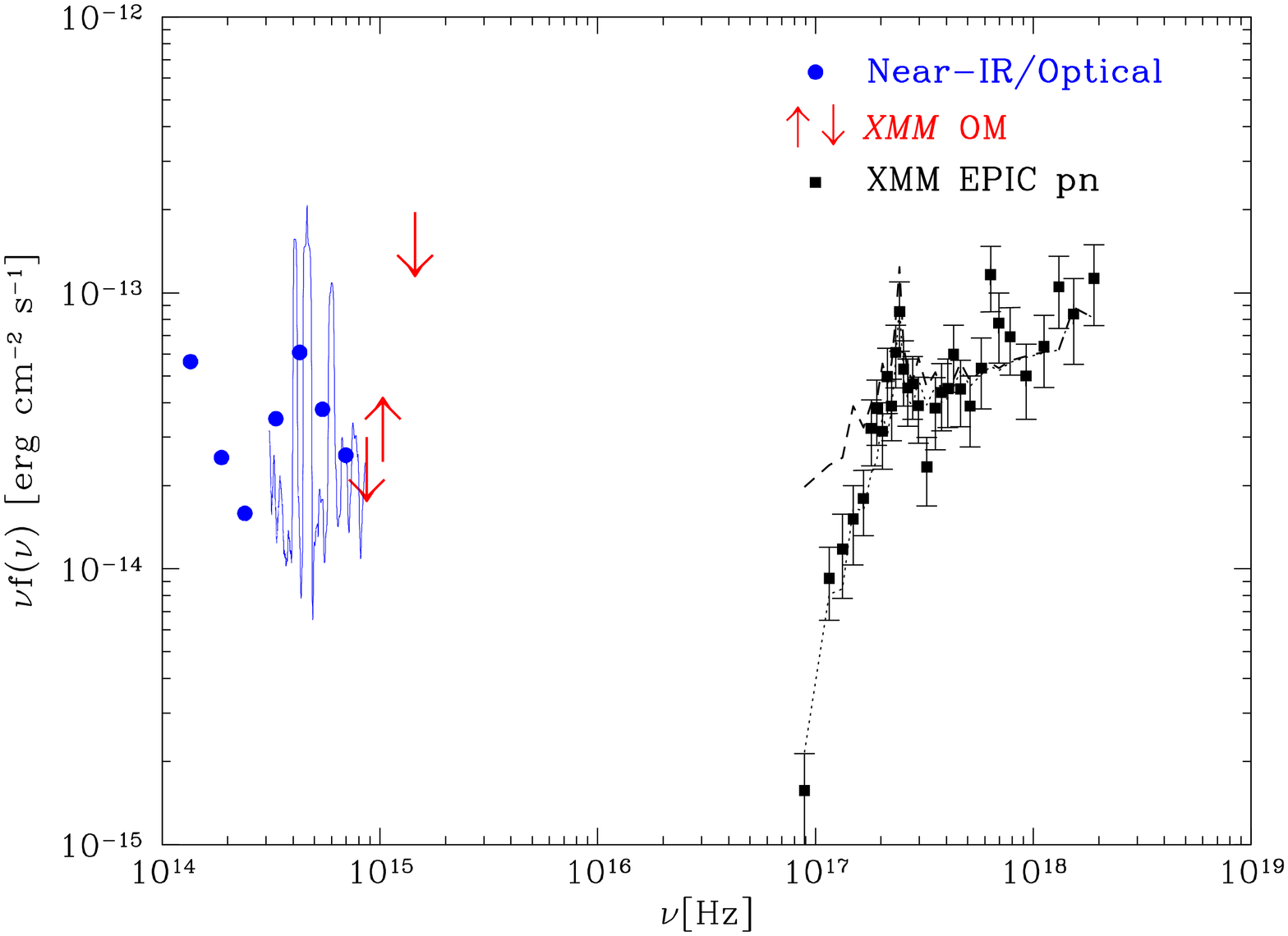} \end{center}
 \caption{Infrared through X-ray spectral energy distribution of SN
 1995N from our coordinated ESO-{\it XMM} observations of July
 2003. The absorbed/unabsorbed X-ray best fitting spectrum is shown as
 a dotted/dashed line. The optical photometry is also shown, superimposed
 on the spectrum (solid line). The arrows represent lower/upper limits for the
 {\it XMM} OM. }
 \label{fig6}
\end{figure*}

\section{Discussion}
\label{sec7}

SN 1995N is one of the few supernovae detected in X-rays at an age of
$\sim$ 9 years (see e.g. SN 1988Z). Until the last {\it ASCA}
observation of January 1998, SN 1995N did not appear to show any
significant decline in the X-ray flux. In fact, on the basis of the
{\it ROSAT} and {\it ASCA} data alone, there was some evidence that
from 2 to 3.5 years after the explosion the X-ray flux may have first
dimmed by 30\% and then brightened \citep{fox00}. 
Our {\it XMM} observation shows that, at an age of
$\sim$ 9 years, almost one order of magnitude decrease in the X-ray
flux of SN 1995N has occurred (see Figure \ref{fig3}).

Using a recent {\it Chandra} observation (March 2004), \cite{chan05}
find that, after subtracting the contamination of nearby sources from
the {\it ASCA} measurement, the light curve in the soft X-ray band
(0.1-2.4) appears to be consistent with a linear decline. However,
considering also the hard X-ray emission, the evolution of the
X-ray light curve appears more complex.
According to \cite{chugai93} and \cite{fox00}, the early behavior of the
X-ray light curve (up to 3.5 years after the explosion) may originate
from a clumpy CSW, where variations of the clump properties
(e.g. density) with distance result in fluctuations of the X-ray
flux. But such variations may also be consistent with a smooth CSW,
where there is some degree of dishomogeneity along the radial
direction. In any case, the significant decline at an age $\sim 9$
years shows that the supernova may have finally entered the X-ray
decline phase. Indeed, from all the available X-ray data, an
interesting scenario emerges where a two-phase CSM contributes to the
X-ray emission. Until $\sim$ 3-4 years, the shock crosses a rather
clumpy wind and the X-ray flux is strongly contaminated by clump
emission, while later the CMS becomes smoother and emission is
dominated by interaction with the diffuse gas.

We find statistically significant evidence for two thermal (MEKAL)
components from the best fit of the {\it XMM} EPIC spectrum. The
temperature of the cooler phase ($kT_1=0.8$ keV),
is consistent with the temperature of the thermal plasma behind the
reverse shock \citep{cf94}. The temperature of the hotter phase
($kT_2=9.5$ keV) is similar to the temperature of the single-component
spectral fit of the {\it ASCA} data performed by \cite{fox00}. Thus
this component, that dominates the high-energy tail of the spectrum,
did not undergo significant spectral evolution, similarly to what
observed for the high-temperature component of the X-ray spectrum of
SN 1978K \citep{schlegel04}. As noted by \cite{fox00}, the temperature
of this hot gas is lower than expected for emission from the forward
shock wave. However, if we interpret it as the temperature of the gas
heated by the forward-shock, we can derive the exponent of the ejecta
density distribution $n$ from the expression $T_1/T_2 =
(3-s)^2/(n-3)^2$, where $s$ is the exponent of the CMS density
distribution \citep{fran96}. Assuming a constant and homogeneous
stellar wind ($s=2$), from the values of the temperatures $T_1$ and
$T_2$ inferred from the X-ray spectrum we obtain $n=6.4$, rather low
but still consistent with the range of values reported by \cite{cf94}
(see also the recently reported value $n\sim 5$ for SN 1978K;
\citealt{schlegel04}).

From the X-ray luminosity of the reverse shock it is possible to
derive an estimate of the mass loss rate of the super-giant progenitor
from which the CSW originated. Assuming that the post-shock gas is
cooling efficiently through line emission, the total luminosity of the
reverse shock is \citep{cf01}
\begin{equation}
L_r = 1.6 \times 10^{41} c_n \frac{{\dot M}_{-5}}{{v_{csw}}_1} V_4^3
\quad {\rm erg} \, {\rm s}^{-1} \, ,
\label{eq:lr}
\end{equation}
where ${\dot M}_{-5}$ is the mass loss rate in units of $10^{-5}
M_\odot \, {\rm yr}^{-1}$, ${v_{csw}}_1$ the CSW velocity in units of
10 km s$^{-1}$, $V_4$ the ejecta velocity close to the shock in units
of $10^4$ km s$^{-1}$, and $c_n=(n-3)(n-4)/(n-2)^3$. From the
unabsorbed flux of the low temperature MEKAL component ($F_{X,l}\simeq
3 \times 10^{-14}$ erg cm$^{-2}$ s$^{-1}$) we derive $L_r=2 \times
10^{39}$ erg s$^{-1}$ ($d=24$ Mpc). Taking $n=6.4$ we then obtain
${\dot M}_{-5}/{v_{csw}}_1 \simeq 0.13 V_4^{-3}$. As reported in the
previous section, the simultaneous VLT spectrum shows that the
velocity of the broader, faster component of the H balmer lines is
$V\simeq 2700$ km s$^{-1}$. Taking this velocity as representative of
the bulk ejecta velocity, we find ${\dot M} \sim 7 \times 10^{-5}
M_\odot \, {\rm yr}^{-1}$ for $v_{csw} \simeq 10 \, {\rm km \,
s}^{-1}$.

At an age of $\sim$ 9 years, the X-ray luminosity of SN 1995N is still
dominated by a high temperature thermal component. If this represents
emission from gas behind the forward shock, we can use also the
luminosity of this component to derive an estimate of ${\dot M}$. In
fact, assuming electron-ion equipartition, the total free-free
luminosity of the forward shock is \citep{cf01}
\begin{equation}
L_f \approx 3 \times 10^{39} {\bar g}_{ff} \left(\frac{{\dot
M}_{-5}}{{v_{csw}}_1}\right)^2 \left(\frac{t}{10 \, {\rm
days}}\right)^{-1} \quad {\rm erg} \, {\rm s}^{-1} \, ,
\label{eq:lf}
\end{equation}
where ${\bar g}_{ff}$ is the free-free Gaunt factor. Because in the
typical physical conditions behind the forward shock electrons and
ions are not in thermal equilibrium, equation (\ref{eq:lf}) is meant
to provide only a crude estimate of ${\dot M}_{-5}/{v_{csw}}_1$. From
the unabsorbed flux of the high temperature MEKAL component
($F_{X,h}\simeq 1.5 \times 10^{-13}$ erg cm$^{-2}$ s$^{-1}$) we derive
$L_f=10^{40}$ erg s$^{-1}$ ($d=24$ Mpc). At $t\sim 3300$ days (epoch
of the {\it XMM} observation), we then have ${\dot M}_{-5}/{v_{csw}}_1
\sim 30$ and hence ${\dot M} \sim 3 \times 10^{-4} M_\odot \, {\rm
yr}^{-1}$. The two estimates of the mass loss rate prior to the
explosion of the progenitor of SN 1995N appear to be roughly
consistent. Averaging between them, we find ${\dot M} \sim 2 \times
10^{-4} M_\odot \, {\rm yr}^{-1}$, at the upper end of the interval
exhibited by red super-giants.

Albeit the smooth wind/ejecta interaction model appears to be
qualitatively consistent with the {\it XMM} data, there are however
some quantitative difficulties. 
The value of the column density of the cool shell is \citep{cf01}
\begin{equation}
N_c \approx 10^{21} (n-4) \frac{{\dot M}_{-5}}{{v_{csw}}_1} V_4^{-1}
\left(\frac{1}{100 \, {\rm days}}\right)^{-1} \quad {\rm cm}^{-2} \, .
\end{equation}
For ${\dot M}_{-5}/{v_{csw}}_1 \sim 20$ and $V_4 \simeq 0.27$, we
obtain $N_c \sim 5 \times 10^{21}$ cm$^{-2}$, significantly larger
than the value of $N_H$ derived from the X-ray spectral fit of the
{\it XMM} data. Furthermore, in the model of \cite{cf94} the forward
shock is assumed to be adiabatic, while the reverse shock is
radiative. The critical time when the cooling time equals the time
since explosion for the forward shock is (\citealt{cf94,cf01})
\begin{equation}
t_c \simeq 1.3 \times 10^4 \left(\frac{n-3}{n-2}\right)
\left(\frac{{\dot M}_{-5}}{{v_{csw}}_1}\right) V_4^{-3} \, s \, .
\label{eq:tc}
\end{equation}
For typical parameters, $t_c \sim 10^4$ s and hence radiative cooling
is not significant for the forward shock (while it may be important
for years for the reverse shock). Using again ${\dot
M}_{-5}/{v_{csw}}_1 \sim 20$ and $V_4 \simeq 0.27$, the critical time
for SN 1995N is $t_c \sim 1$ year. Radiative cooling may still be
significant at the time of the {\it XMM} observation. Assuming that
the forward shock emission dominates, $L_f \propto t^{-1}$ and the
forward shock luminosity at the epoch of the {\it ASCA} observation
would have been 2.5 times larger, which is not sufficient to account
for 0.2-10 keV flux inferred from the {\it ASCA} measurement, even taking
into account the contamination from nearby sources.
This may be further evidence that the emission up to 3.5 years was
contaminated by clumps.
It should be noted, however, that a certain spectral evolution at high
energies between the {\it ASCA} and {\it XMM} observations would be
expected, whereas the observations do not show any evidence of it.

Because of the decrease in the X-ray flux and the strong contamination
by solar flares, the statistics of the EPIC spectrum is not sufficient
to detect X-ray spectral lines. Therefore, we can not test the
suggestion by \cite{fox00} about the possible existence of a Silicon
spectral feature near 1.8 keV, although some residuals are present in
the EPIC pn spectrum at 1.8, 2.6 and 5.4 keV. However decline in the
X-ray flux is signaling that the SN has started to evolve towards the
remnant stage, in agreement also with the increase in strength of the
forbidden O lines observed in the optical spectrum
\citep{pasto05}. Future X-ray observations will then be important for
detecting X-ray line emission. In this respect, we note that, in a
{\it Chandra} spectrum taken almost 9 months later, \cite{chan05}
report evidence of emission features around 1 keV that they identify
as Ne lines.

Thanks to our coordinated {\it XMM}-ESO VLT observing program we have
been able to construct the simultaneous broadband flux distribution of
SN 1995N from the IR to the X-ray bands (see Table \ref{tab3} and
Figure \ref{fig6}). Most of the total energy output comes from the
direct X-ray thermal emission due to the wind/ejecta interaction,
whereas, at $\sim$ 9 years after the explosion, reprocessed optical
and near IR emission are less important. The $K$ band emission of SN
1995N is believed to be produced by reprocessing of X-ray radiation by
dust formed in the ejecta or pre-existing in the CSM (see
e.g. \citealt{gerardy02}). Together, the fluxes in the $R$ and $K$
bands amount to $\approx$ 15\% of the X-ray flux.
The $R$ band flux is approximately 2 times larger than that in the
neighbouring optical bands because of the significant contribution of
a few prominent lines (H$\alpha$, [OI] 6300-6364, [OII] 7320-7330). In
particular, the H$\alpha$ line is thought to originate from
recombination and collisional excitation of gas in the photo-ionized
ejecta and in the cool shell bound by the reverse shock and the
contact discontinuity. The H$\alpha$ flux measured from the optical
spectrum is 5.5 $\times$ 10$^{-15}$ erg cm$^2$ s$^{-1}$.
This is $\sim$ 20\% of the flux of the low
temperature component. This result does not appear to be in agreement
with the findings of \cite{cf94} that $\sim$ 1 \% of the reverse shock
luminosity is emitted as H$\alpha$, fairly independent of density and
other physical parameters. However, it should be noted that
significant reverse shock emission may be present in the far-UV, soft
X-ray (below 0.2 keV) spectral region, thus increasing the total X-ray
flux.

The detection of SN 1995N in the $UVW1$ band of the {\it XMM} OM shows
that significant UV emission accompanies the X-ray luminosity, in
agreement with the reprocessing of the X-ray flux that is expected to
occur in these environments. The origin of this emission is usually
ascribed to Ly$\alpha$ and other lines of highly ionized species, as C
III-IV, NIII-V, O III-VI, that are produced in the un-shocked ejecta,
photo-ionized by the reverse shock (e.g. \citealt{cf01}).

At the epoch of our multi-wavelength observation, the IR emission has
greatly diminished. The power emitted in the $K$ band amounts to
$\sim$ 7\% of the total X-ray flux. This late time near IR luminosity
is consistent with the simple picture of re-processing of a small
fraction of the optical and X-ray radiation by a pre-existing dusty
circumstellar medium heated by the supernova shock (see
e.g. \citealt{gerardy02}). As the shock moves away from the region
where the bulk of the dust is located, the emission quenches.

\section{Conclusions}
\label{sec8}

We observed SN 1995N with {\it XMM}-Newton at $\sim$ 9 years after the
estimated date of explosion. We found that the 0.2--10.0 keV flux of
SN 1995N dropped at a value of $\simeq 1.8 \times 10^{-13}$ erg
cm$^{-2}$ s$^{-1}$, about one order of magnitude lower than that of a
previous {\it ASCA} observation performed more than 6 years ago
(January 1998). The decline in the X-ray flux is signaling that SN
1995N has probably started to evolve towards the remnant stage.

Interpreting the evolution of the X-ray light curve is not
straightforward. A complex scenario where a two-phase (clumpy and
smooth) CSM contributes to the observed X-ray emission is consistent
with the available data.

The EPIC spectrum of SN 1995N shows statistically significant evidence
for the presence of two distinct thermal (MEKAL) components at
temperatures of 0.8 and 9.5 keV, respectively. If they are interpreted
as the temperatures of the gas behind the reverse/forward shocks, we
derive that the exponent of the ejecta density distribution is
$n=6.4$.

Assuming that the observed X-ray luminosities of the hard/soft
spectral components represent the forward/reverse shock emission, we
derive an estimate of the mass loss rate of the progenitor, ${\dot M}
\sim 2 \times 10^{-4} M_\odot \, {\rm yr}^{-1}$, at the upper end of
the interval exhibited by red super-giants.

The X-ray data were obtained within the framework of a multi-wavelength
(IR/optical/UV/X-ray) observational campaign. The simultaneous
infrared to X-ray flux distribution of SN 1995N shows that the
direct X-ray thermal emission due to the wind/ejecta interaction
is $\sim 5$ times larger than the total reprocessed IR/optical flux.
At the epoch of our multi-wavelength observation, the IR emission
has greatly diminished.

\section*{Acknowledgments}

This work has been partially supported by the Italian Ministry for
Education, University and Research (MIUR) under grant
PRIN-2002-027145. We thank an anonymous referee for valuable comments.

\bsp

\label{lastpage}

\end{document}